\documentclass[conference]{IEEEtran}
\usepackage{cite}
\usepackage{amsmath,amssymb,amsfonts}
\usepackage{textcomp}

\usepackage{enumitem}
\usepackage{listings}
\usepackage{float}
\usepackage{algorithm}
\usepackage[noend]{algpseudocode}
\usepackage{tabularx}
\usepackage[caption=false,font=footnotesize]{subfig}
\usepackage{graphicx}
\usepackage{placeins}
\usepackage{wrapfig}
\usepackage{booktabs}

\usepackage[hidelinks]{hyperref}
\newcommand{\secref}[1]{Section~\ref{#1}}

\def\BibTeX{{\rm B\kern-.05em{\sc i\kern-.025em b}\kern-.08em
    T\kern-.1667em\lower.7ex\hbox{E}\kern-.125emX}}

\begin{document}

\def\system{MARS}

\def\longname{Multipath Adaptive Reliable Service}

\title{MARS: Multipath Adaptive Reliable Service}

\author{
\IEEEauthorblockN{Yitong Li, Xinjiao Li, and Dirk Kutscher}
\IEEEauthorblockA{The Hong Kong University of Science and Technology (Guangzhou)\\
\texttt{\{yliop,xli886\}@connect.hkust-gz.edu.cn}, \texttt{dku@hkust-gz.edu.cn}}
}

\hypersetup{
    pdftitle={MARS: Multipath Adaptive Reliable Service},
    pdfauthor={Yitong Li, Xinjiao Li, Dirk Kutscher}
}

\maketitle

\begin{abstract}
Multipath transport is important for Internet/WAN services that move large data volumes across heterogeneous paths, including geo-distributed analytics, content distribution, and cloud-service pipelines. Existing solutions face a trade-off: end-to-end transports such as MPTCP and MPQUIC are deployable but limited by endpoint-visible paths and delayed congestion feedback, while routing- or forwarder-assisted approaches often require infrastructure support or lack safe coordination across forwarding choices.

This paper presents \system{}, a receiver-driven, forwarder-assisted multipath transport for Internet/WAN environments. \system{} combines tier-synchronized overlay path discovery with coupled consumer/forwarder congestion control, enabling it to safely expand usable forwarding opportunities and react near bottlenecks. It runs as an incrementally deployable UDP overlay at clients, servers, relays, or CDN-like nodes. We implement \system{} in simulation and as a prototype, and evaluate it through simulation and Mininet emulation across deployment scopes, loss rates, and a forwarding-face outage scenario.

Results show \system{} provides deployment-dependent benefits: with endpoint-only deployment, it performs comparably to the evaluated ECMP-limited configurations of MPTCP and MPQUIC. With cooperating overlay forwarders, it safely expands the usable path set from routing-exposed forwarding candidates. Across the tested loss conditions, it reduces maximum $T_{95}$ by up to 66.7\% and 63.9\% relative to the evaluated path-expanded MPTCP and MPQUIC configurations, respectively, given the same path set. Path discovery remains lightweight, flow fairness remains high, and \system{} degrades gracefully during an emulated forwarding-face outage and recovers quickly after face restoration. Overall, ICN-style receiver-driven forwarding can serve as a deployable overlay transport substrate for coordinated WAN multipath without requiring changes to IP routing.

\end{abstract}

\section{Introduction}
\label{sec:intro}

Modern Internet/WAN services increasingly move large data volumes across sites. Global traffic continues to grow, while geo-distributed AI, recommender, and cloud-service pipelines generate bandwidth-intensive exchanges~\cite{itu2024traffic,ericsson2025mobility,atlas2024geolm,netstorm2024geodml,youtubeDNN2016}. A single WAN path may be congested, lossy, or rate-limited even when other endpoint- or relay-accessible paths retain capacity. Multipath transport can therefore improve aggregate throughput, resilience, and performance stability~\cite{rfc8684}.

Existing approaches expose path diversity at different layers. End-to-end transports such as CMT-SCTP, MPTCP, and MPQUIC are incrementally deployable but limited to endpoint-visible addresses, interfaces, or path identifiers and cannot coordinate near internal bottlenecks~\cite{cmt-sctp,rfc8684,mpquic-paper,mpQUIC}. Routing-level and path-aware schemes, including ECMP and interdomain multipath, can expose richer choices but require router, protocol, or operator support~\cite{ECMP,yamr,miro,rfc9049,schmitt2018pathaware}. Overlays and ICN-style forwarders add intermediate choices above IP~\cite{ICN,ron,overlaytcp,pcon,ndnqsf,mircc}, yet often take routing-provided next hops as given or do not jointly coordinate endpoint scheduling, congestion response, and forwarding. This leaves a tradeoff among deployability, path visibility, and coordination.

Service-managed relays and edge platforms, together with UDP-based user-space transport, make an overlay bridge practical without router changes~\cite{quicRFC}. ICN is a useful substrate because its receiver-driven exchange and stateful forwarding let intermediate nodes participate in forwarding and congestion decisions. However, existing ICN multipath still depends largely on routing-installed faces~\cite{ndndv,nlsr}; using them without coordination can reduce forwarding efficiency. The missing bridge is a deployable overlay that safely expands usable forwarding opportunities while coordinating consumers and forwarders.

We present \system{}, a multipath transport framework for Internet/WAN environments that bridges endpoint-only and routing-level multipath through tier-synchronized overlay path discovery and coupled consumer/forwarder congestion control. It safely expands the usable end-to-end path set from routing-exposed forwarding candidates and reacts closer to WAN bottlenecks. As an incrementally deployable UDP overlay, \system{} requires no changes to IP routers or routing protocols.

This paper makes the following contributions:
\begin{itemize}[leftmargin=*]
    \item We identify a deployable multipath gap in Internet/WAN environments: existing approaches either remain limited by endpoint- or routing-visible paths, require routing-plane or operator support, or lack unified transport-level coordination across path discovery, congestion response, and forwarding.
    
    \item We design a tier-synchronized overlay path discovery mechanism that expands usable forwarding opportunities beyond routing-exposed paths while preventing looped Interests, backward steering, and excessive path inflation.
    
    \item We design a coupled per-prefix/per-face congestion-control mechanism that uses upstream Interest/Data queue feedback to coordinate consumer request generation with forwarder-side scheduling near bottlenecks.
    
    \item We implement \system{} in simulation and as a working prototype, and evaluate it through large-scale simulation and Mininet emulation under homogeneous and heterogeneous WAN settings. The results show that \system{} remains competitive with endpoint-only deployment, gains substantially from cooperating overlay forwarders, and remains robust under loss and the evaluated forwarding-face outage.

\end{itemize}

\section{Motivation}
\label{sec:motivation}

\begin{figure}[t]
    \centering
    \subfloat[Completion-time degradation\label{fig:pcon-performance}]{%
    \begin{minipage}[c]{0.54\columnwidth}
        \centering
        \includegraphics[width=\linewidth,trim={2 2 2 2},clip]{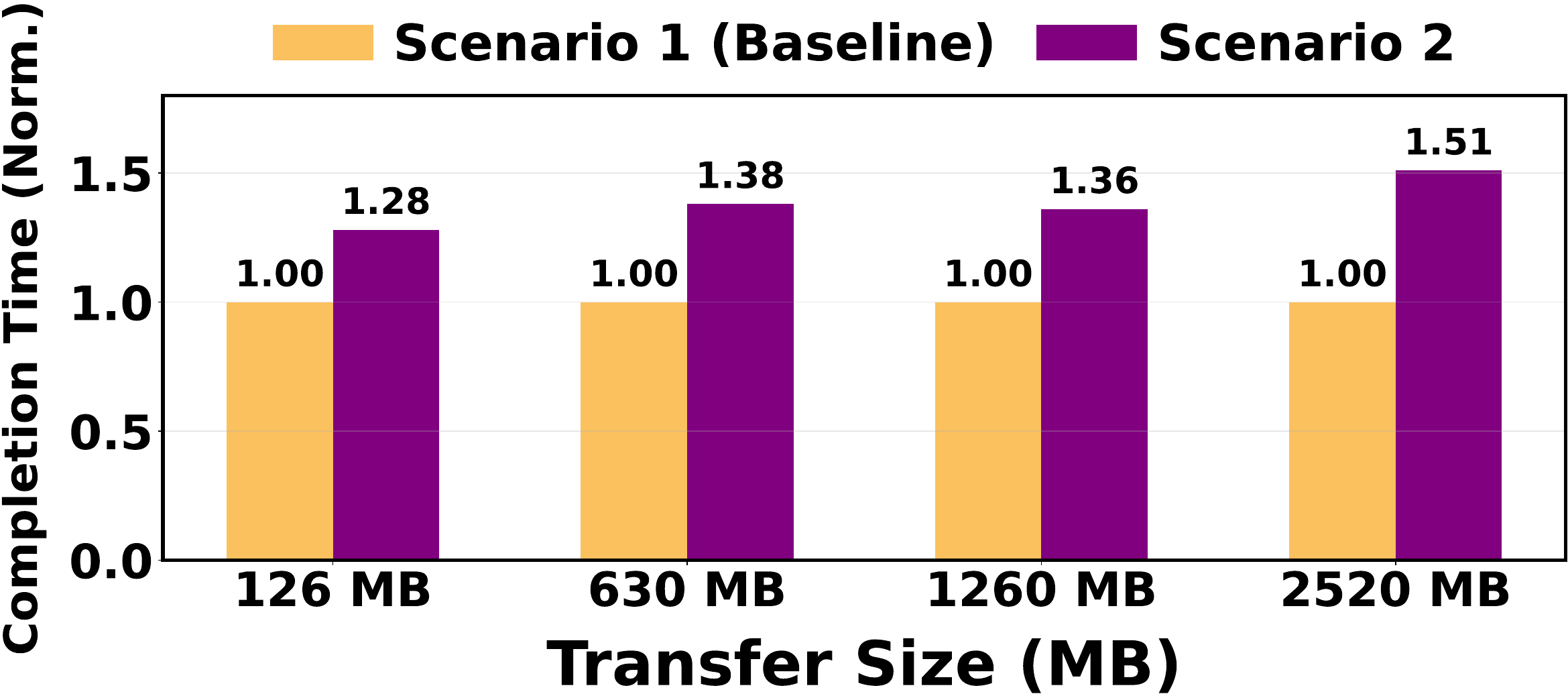}
    \end{minipage}
    }
    \hfill
    \subfloat[Forwarding scenarios\label{fig:pcon-loop-scenarios}]{%
    \begin{minipage}[c]{0.43\columnwidth}
        \centering

        \makebox[\linewidth][c]{\footnotesize\textbf{Scenario 1: loop-free}}\\[0.15em]
        \includegraphics[width=0.90\linewidth,trim={2 2 2 2},clip]{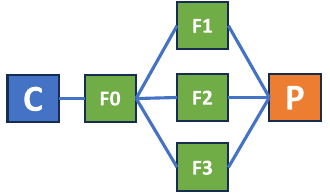}

        \vspace{0.45em}

        \makebox[\linewidth][c]{\footnotesize\textbf{Scenario 2: loop-prone}}\\[0.15em]
        \includegraphics[width=0.90\linewidth,trim={2 2 2 2},clip]{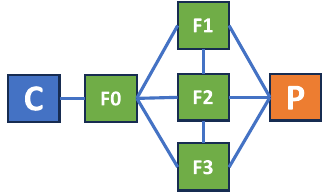}
    \end{minipage}
    }

    \caption{PCON~\cite{pcon} performance degradation under loop-free and loop-prone forwarding scenarios.}
    \label{fig:pcon-motivation}
\end{figure}

\textbf{\textit{End-to-end multipath is bounded by routing-visible paths.}}
End-to-end multipath transports such as MPTCP~\cite{mptcp} and MPQUIC~\cite{mpquic-paper} do not create new forwarding opportunities by themselves; they schedule traffic over paths exposed through endpoint addresses, subflows, interfaces, or transport-level path identifiers~\cite{rfc8684,mpQUIC}. When multiple subflows share the same endpoint pair, path separation depends on the underlying routing or load-balancing system, such as ECMP hashing across subflows~\cite{raiciu2011mptcpdc}. However, ECMP only distributes traffic among equal-cost next hops using flow-level hashing~\cite{ECMP}, while Internet paths are often asymmetric and policy-dependent~\cite{paxson1996endtoend,devries2015asymmetric}. As a result, endpoint-only multipath may expose multiple subflows at the connection layer while still exercising only a small set of useful WAN forwarding opportunities.

\textbf{\textit{Routing-level multipath depends on infrastructure support.}}
Routing-level and path-aware approaches expose richer path choices through routers, routing protocols, or network domains, but this visibility comes at the cost of infrastructure support. Maintaining multiple policy-compliant paths requires routing-plane changes, operator coordination, or path-aware mechanisms that are difficult to deploy incrementally across independent Internet/WAN domains~\cite{yamr,miro,rfc9049,schmitt2018pathaware}. Moreover, these mechanisms operate mainly through routing or control-plane decisions, rather than being tightly coupled with per-flow congestion control, queue dynamics, or transport-layer scheduling. Thus, they improve path visibility but do not provide a practical transport-level mechanism for fast congestion response and coordinated load balancing.

\textbf{\textit{Forwarder-assisted multipath lacks safe path expansion.}}
Overlay- and forwarder-assisted systems improve deployability by introducing relays, proxies, or named-data forwarders above IP routing~\cite{ron,overlaytcp}. ICN is a natural substrate for this direction because its receiver-driven Interest/Data exchange and stateful per-prefix forwarding allow intermediate forwarders to adapt upstream choices without changing IP routers. Existing ICN-based designs further use congestion signals to guide Interest forwarding~\cite{pcon,mircc,ndnqsf}. However, their forwarding choices still largely come from ICN routing protocols, which may install multiple next hops beyond a single shortest path~\cite{routing-muca,nlsr,ndndv}; these choices are not automatically safe or efficient. In Internet/WAN overlays, blindly using all available next hops can lead to \textit{looped Interests, backward steering, and path inflation}, where requests cycle among forwarders, move away from the producer, or traverse unnecessarily long paths.

Our preliminary experiment in Fig.~\ref{fig:pcon-performance} confirms the impact of this problem under the two scenarios in Fig.~\ref{fig:pcon-loop-scenarios}. Completion time is normalized by the loop-free baseline for each transfer size, highlighting the relative performance loss caused by unsafe forwarding choices. Although the \textit{loop-prone topology} is intentionally simple, it isolates the effect of unsafe next-hop expansion; we revisit this effect under larger concurrent WAN transfers in \secref{sec:evaluation}. Across different transfer sizes, PCON's completion time increases by 28\%--51\%, and the degradation becomes worse as the transferred data volume grows. This shows that, without coordination, forwarding loops can persistently waste bandwidth and severely reduce efficiency for large-volume transfers.

These limitations suggest three key requirements for a practical Internet/WAN multipath transport. 
\textbf{First, incremental deployability and infrastructure compatibility:} the system should operate over today’s Internet/WAN infrastructure without requiring changes to IP routers, routing protocols, or interdomain control planes, while remaining flexible enough to be deployed through end hosts, overlays, or selected forwarders. 
\textbf{Second, richer path visibility with safe path expansion:} the system should expose useful forwarding opportunities beyond endpoint-visible paths and ECMP-selected routes, but must preserve forwarding efficiency by avoiding looped Interests, backward steering, and excessive path inflation. 
\textbf{Third, unified coordination of multipath forwarding and congestion control:} the system should coordinate path discovery, load balancing, congestion response, and reliability under one transport-level abstraction, so that additional path diversity improves performance rather than introducing uncoordinated forwarding anomalies.

\section{\system~ Design}
\label{sec:\system{}}

\begin{figure}[t]
    \centering
    \includegraphics[width=\columnwidth]{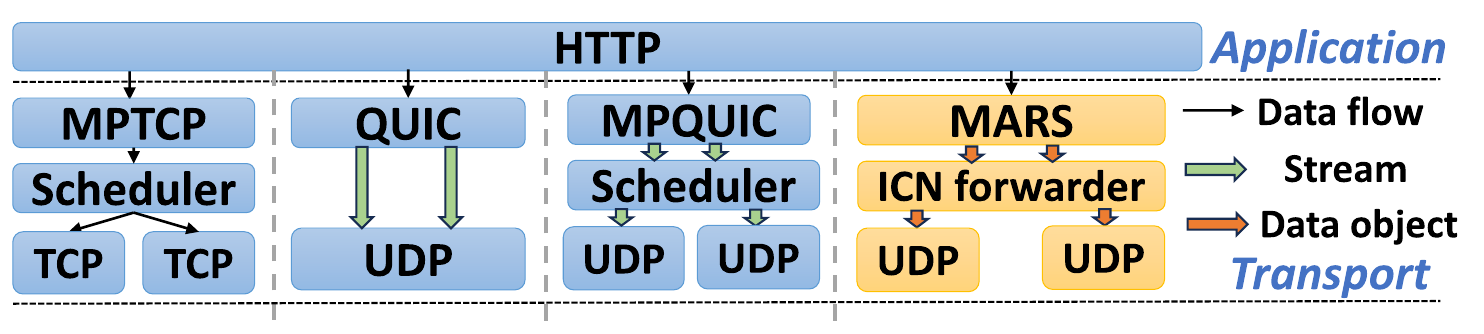}
    \caption{\system{} vs. endpoint-only multipath transports.}
    \label{fig:high-level-architecture}
\end{figure}

\begin{figure}[ht]
  \centering
  \includegraphics[width=0.98\linewidth]{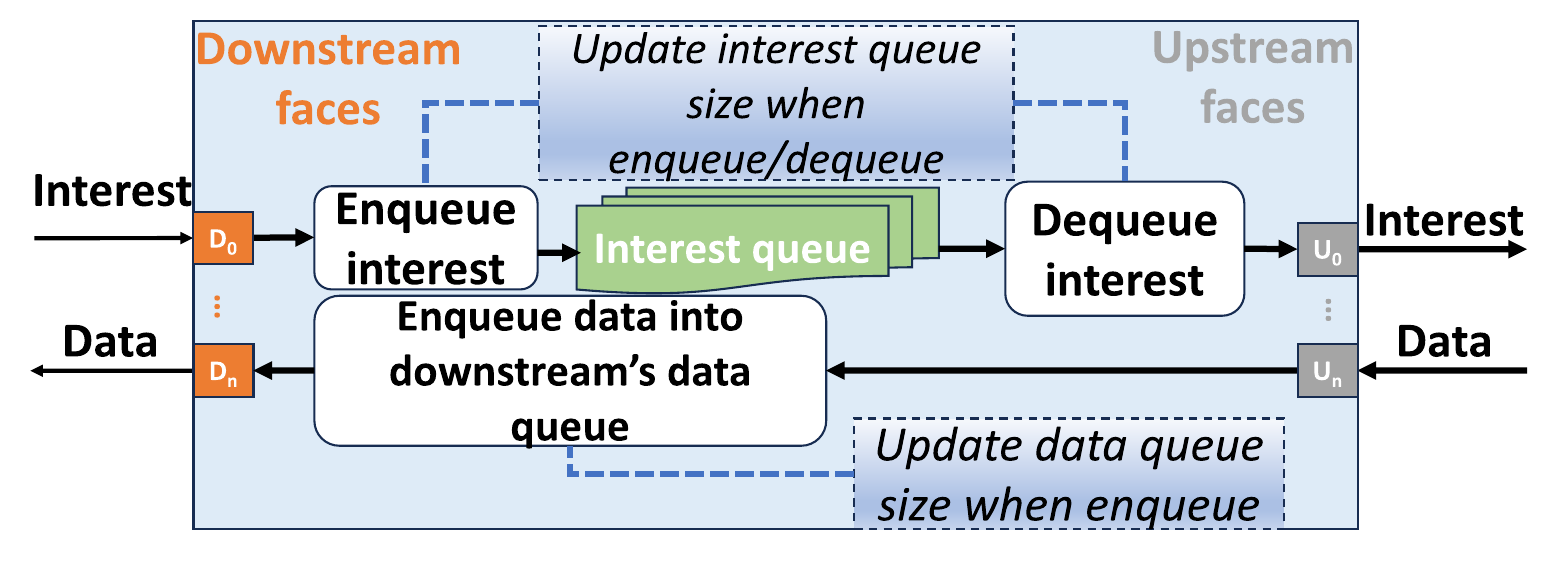}
  \caption{\system~ forwarder design.}
  \label{fig:forwarder}
\end{figure}

\system{} is a receiver-driven, forwarder-assisted multipath transport built over an ICN-style UDP overlay. Consumers retrieve named objects through Interests, while intermediate \system{} forwarders maintain per-prefix queues, schedule Interests over upstream faces, and return Data along reverse paths; producers remain transparent to the multipath logic. As Fig.~\ref{fig:high-level-architecture} illustrates, this adds lightweight overlay coordination beyond endpoint-only MPTCP/MPQUIC and supports deployment at selected service nodes without changing IP routing.

\textbf{Trust boundary.}
\system{} targets service-managed or cooperative overlays: participating forwarders are authorized by the service operator or cooperating entities and assumed non-Byzantine; they are trusted to forward protocol messages and report queue metadata faithfully. Hop-by-hop authentication and integrity can protect \system{} messages and metadata from external on-path modification, but cannot prevent an authorized forwarder from reporting false state. Defending against such behavior would require attestation or Byzantine-resilient control and is outside our threat model and evaluated claims.

\subsection{Architecture Overview and Forwarding Substrate}
\label{subsec:design-overview}

The \system{} forwarding substrate provides the state and signals required by the two core algorithms. Each forwarder maintains upstream and downstream faces, per-prefix Interest queues for shaping outgoing requests, and queue-size feedback for tracking upstream congestion. Its key design choice is to regulate traffic only in the Interest direction: Interests are shaped before being forwarded upstream, while Data packets follow the normal downstream forwarding path and carry lightweight queue feedback as metadata. This allows downstream forwarders and consumers to observe upstream queue conditions and react closer to where congestion appears. \system{} also uses separate naming formats for path discovery and data transmission, so that probing state and normal retrieval traffic can be processed independently.

\subsubsection{Queues}
\label{subsubsec:queues}

Each forwarder maintains two types of queues. First, it maintains a \textit{per-prefix Interest queue} with finite capacity\footnote{The effective per-prefix Interest-queue capacity is 40 packets in our two-thread prototype.}. This queue is specific to \system{} and is used to shape the Interest forwarding rate for each requested prefix. When an Interest is admitted into the queue, the forwarder can regulate when it is released to an upstream face.

Second, each downstream face maintains a standard \textit{Data queue} for serializing outgoing Data packets. \system{} does not shape this queue: once a Data packet arrives from an upstream face, it is forwarded downstream according to normal face scheduling. The Data queue is therefore not a separate traffic control mechanism, but it is still monitored because its occupancy reflects downstream congestion.

\subsubsection{Queue-derived Congestion Feedback}
\label{subsubsec:queue-signal}

\system{} exposes congestion through normalized queue-occupancy feedback carried in Data packet metadata. For an upstream face $f$, $q_{face}^{int}(f)$ and $q_{face}^{data}(f)$ denote the Interest- and Data-queue occupancies reported by the upstream forwarder connected through $f$. Each occupancy is normalized by its own queue capacity, so both signals lie in $[0,1]$ and can be compared directly. These values therefore represent the congestion state of the upstream forwarding choice, not the local queue occupancy of the downstream node. Each forwarder updates Data metadata with its latest normalized occupancies before forwarding the packet downstream. This allows downstream forwarders and consumers to observe upstream congestion without explicit control messages, and the congestion control algorithm uses these feedback signals and their temporal changes to adjust Interest sending rates.

\begin{figure}[ht]
    \centering
    \begin{minipage}[t]{0.48\columnwidth}
        \centering
        \includegraphics[width=\linewidth]{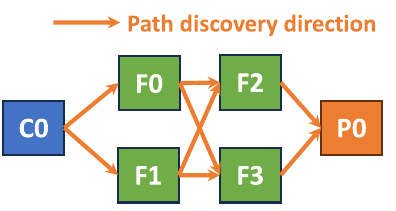}
        \caption{Path discovery example.}
        \label{fig:demo-pd}
    \end{minipage}
    \hfill
    \begin{minipage}[t]{0.48\columnwidth}
        \centering
        \includegraphics[width=\linewidth]{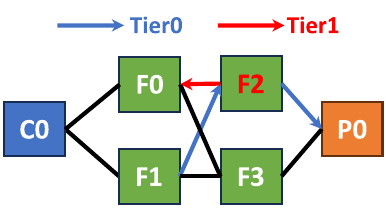}
        \caption{Challenge without synchronization in path discovery.}
        \label{fig:challenge-without-sync}
    \end{minipage}
\end{figure}

\begin{figure*}[t]
  \centering
  \includegraphics[width=0.9\textwidth]{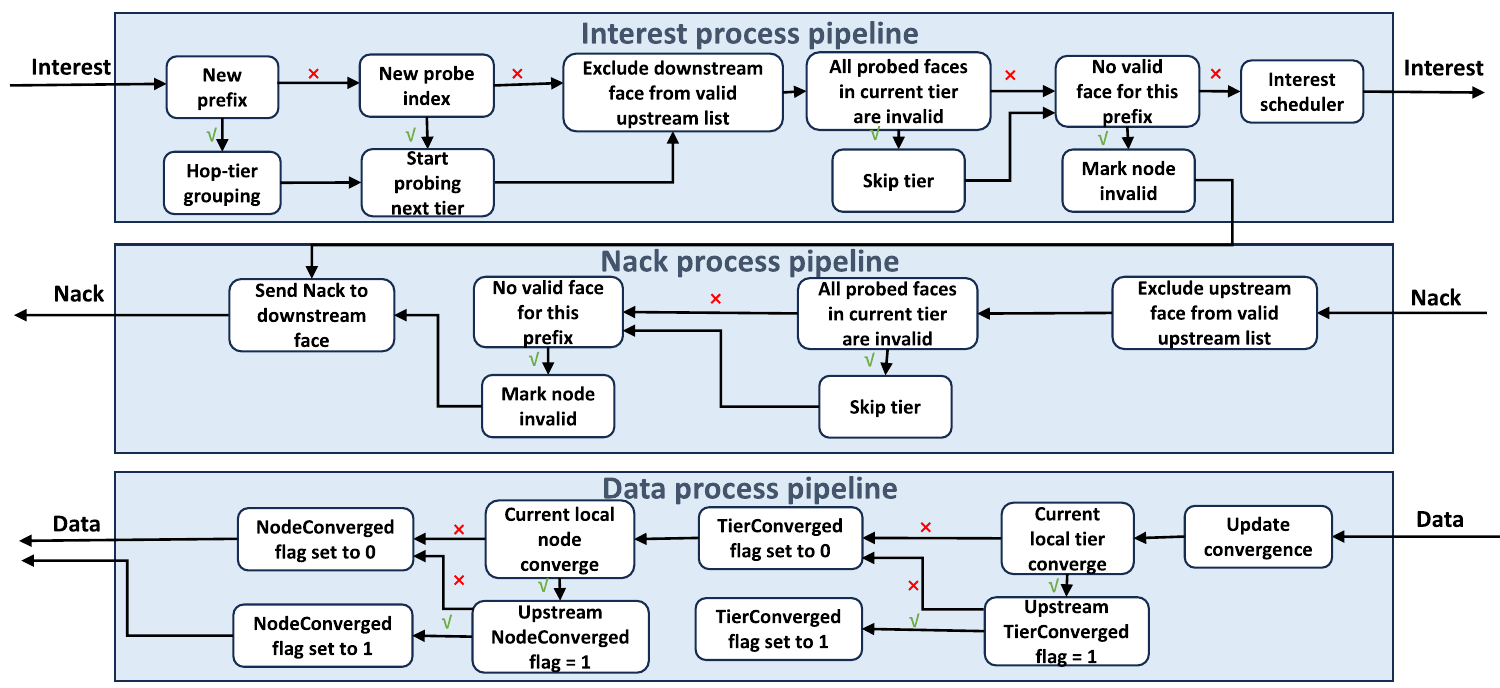}
  \caption{Path discovery pipeline.}
  \label{fig:PathDiscoveryPipeline}
\end{figure*}

\subsubsection{Hierarchical Naming Specification}
\label{subsec:naming_spec}

\system{} defines an ICN-style hierarchical naming scheme to distinguish path-discovery probes from normal data retrieval. We use two name formats:

\begin{itemize}[leftmargin=*]
    \item \textbf{Path Discovery Format:} To facilitate the route exploration described in \secref{subsec:path-discovery}, we define a five-component hierarchy: \texttt{/dest/src/pd/tier/seq-N}. Here, \texttt{dest} represents the target producer, \texttt{src} represents the requesting consumer, and \texttt{pd} identifies a path-discovery message.
    The \texttt{tier} component is the consumer-controlled discovery index used to synchronize tier advancement, and \texttt{seq-N} carries the sequence number.
    
    \item \textbf{Data Transmission Format:} For standard content retrieval and congestion-control feedback, we use a four-component hierarchy: \texttt{/dest/src/dt/seq-N}. The \texttt{dest} component is the routable producer prefix, \texttt{src} identifies the requesting consumer, \texttt{dt} denotes data transmission, and \texttt{seq-N} identifies the chunk sequence.
\end{itemize}

\subsection{Tier-Synchronized Overlay Path Discovery}
\label{subsec:path-discovery}

To enable safe path expansion, our design builds path discovery into a tier-based probing process (see \secref{subsubsec:hop-tier-grouping}). To avoid unsafe tier advancement without global synchronization, path discovery is coordinated in a consumer-driven manner
(see \secref{subsubsec:synchronization}). To ensure path discovery converges, efficient convergence mechanisms are implemented (see \secref{subsubsec:converge}). The forwarding logic for Interests, NACKs, and Data is illustrated in Fig.~\ref{fig:PathDiscoveryPipeline}.

\subsubsection{Hop-tier Grouping}
\label{subsubsec:hop-tier-grouping}

Hop-tier grouping takes the upstream faces exposed by the local forwarding information base (FIB) for a given prefix as forwarding candidates and organizes them into tiers according to their routing-provided hop counts. At each round, upstream faces are probed in order of increasing hop count, always starting from the shortest available paths. By grouping probes into hop-tiers and ensuring that a face used as ingress for a shorter path is never selected as an upstream candidate for longer-hop probes, the system avoids backward steering and forwarding loops within the discovered forwarding set.
Ideally, as illustrated in Fig.~\ref{fig:demo-pd}, path discovery should proceed step by step from the source to the destination, following the forward direction without any backward steering, thus enabling scalable discovery of multiple viable forwarding paths.
To further enhance efficiency, we introduce a hop-count threshold: only nexthops satisfying a configured threshold relative to the minimum hop count are considered during path discovery.

\subsubsection{Consumer-Driven Synchronization}
\label{subsubsec:synchronization}

Without synchronization, an upstream forwarder closer to the producer may receive Data before downstream nodes converge and prematurely probe a later tier, potentially steering Interests backward and creating loops, as illustrated in Fig.~\ref{fig:challenge-without-sync}. To prevent this, \system{} uses consumer-driven synchronization instead of a global controller: the consumer embeds a $\mathsf{ProbeIdx}$ field in Interest names and advances it only after all nodes in the current tier have converged, as indicated by the $\mathsf{TierConv}$ flag in Data metadata. This ensures that downstream nodes always begin probing a tier before upstream nodes, preserving forward progress. The $\mathsf{NodeConv}$ flag indicates when path discovery for a prefix has fully converged and serves as the stop criterion. In essence, the consumer coordinates global tier advancement, while each forwarder manages local discovery state and advances only when directed by an updated $\mathsf{ProbeIdx}$ from downstream.

\subsubsection{Efficient Convergence Mechanisms}
\label{subsubsec:converge}

A node updates its convergence status whenever Data arrives from an upstream face, removing that face from \textit{PendingConverged}. \textbf{Tier convergence} occurs once all faces in the current tier have been probed and the downstream metadata flag $Tier_{\text{conv}}{=}1$; \textbf{node convergence} occurs once all tiers have converged and $Node_{\text{conv}}{=}1$. These flags are embedded in Data packet metadata and propagated downstream (only on path discovery).

During path discovery, tier convergence marks completion of probing within a tier, while node convergence indicates that local discovery for the prefix is complete. New Interests are then forwarded round-robin across remaining faces in \textit{ValidUpstream}, ensuring that downstream discovery proceeds correctly. To accelerate convergence, we apply a \emph{tier-skipping} rule: if all faces in the current tier are ineligible, the node immediately declares convergence, since longer-hop tiers are unlikely to yield better paths. 

Path discovery preserves forward progress through three local constraints: (i) a face used as downstream ingress for a shorter-tier probe is not reused as a longer-tier upstream face; (ii) $\mathsf{ProbeIdx}$ lets upstream forwarders probe the next tier only after downstream convergence on the current tier; and (iii) the hop-count threshold limits candidate next hops relative to the shortest known distance. These constraints avoid backward steering and bound path inflation, but do not aim to find globally optimal or maximally disjoint paths.

\subsection{Coupled Consumer/Forwarder Congestion Control}
\label{subsec:congestion-control}

\system{} couples per-prefix consumer rate control with per-face forwarder rate control over the upstream faces selected by path discovery.
Inspired by prior work that uses upstream feedback as an early congestion signal~\cite{ndnqsf}, we treat the normalized upstream Interest- and Data-queue occupancies as indicators of whether a forwarding choice is becoming congested before persistent loss or timeout occurs. Unlike endpoint-only rate control, \system{} applies coordinated rate adaptation at both consumers and forwarders: consumers regulate flow-level Interest generation, while forwarders regulate per-face Interest forwarding over the discovered path set. This unified control loop allows downstream nodes to react to congestion closer to where it appears and to shift load across valid upstream faces in a coordinated manner.

Algorithm~\ref{alg:congestion-control} summarizes the primary controller parameters, which are not tuned per loss condition. $q_{\mathrm{low}}$/$q_{\mathrm{high}}$ set normalized queue-pressure thresholds, $M_D$/$G_D$ control rate decreases, $C_I$/$A_I$ control probing increases, and $\gamma_{\max}$ caps the updated rate once the bandwidth guard is active, while $\gamma_{\min}$ applies only under low queue pressure. Consumers and forwarders use $\gamma_{\max}=1.2$ and $1.3$, respectively. In the evaluated configuration, $\tau_R=20$\,ms, approximately half the observed base RTT.

\begin{algorithm}[t]
\caption{\system{} Congestion Control with Bandwidth Estimation}
\label{alg:congestion-control}
\small
\begin{algorithmic}[1]
\Require Normalized upstream queue feedback $q^{\mathrm{int}}_{\mathrm{face}}(f),q^{\mathrm{data}}_{\mathrm{face}}(f)\in[0,1]$
\Require Queue slopes $s^{\mathrm{int}}_{\mathrm{face}}(f)$, $s^{\mathrm{data}}_{\mathrm{face}}(f)$
\Require Current rate $R$, bandwidth estimate $\hat{B}$, measured throughput $B_m$, throughput trend $T$
\Require $q_{\mathrm{high}}=0.8$, $q_{\mathrm{low}}=0.5$, $s_{\mathrm{thres}}=0.1$
\Require $M_D=0.9$, $G_D=0.95$, $C_I=1.02$, $A_I=1.05$
\Require $\alpha_{\mathrm{strong}}=0.25$, $\alpha_{\mathrm{small}}=0.1$, $\gamma_{\min}=0.5$
\Require Observed base RTT $\mathrm{RTT}_{\mathrm{base}}$; $\tau_B=\mathrm{RTT}_{\mathrm{base}}/6$, $\tau_R=\mathrm{RTT}_{\mathrm{base}}/2$

\Statex
\State \textbf{Select bottleneck queue feedback:}
\If{$q^{\mathrm{int}}_{\mathrm{face}}(f) > q^{\mathrm{data}}_{\mathrm{face}}(f)$}
    \State $q \gets q^{\mathrm{int}}_{\mathrm{face}}(f)$; $s \gets s^{\mathrm{int}}_{\mathrm{face}}(f)$
\Else
    \State $q \gets q^{\mathrm{data}}_{\mathrm{face}}(f)$; $s \gets s^{\mathrm{data}}_{\mathrm{face}}(f)$
\EndIf

\Statex
\State \textbf{Update bandwidth estimate:}
\If{BW update interval $\tau_B$ has elapsed}
    \State $\alpha \gets 0$
    \If{$B_m > \hat{B}$}
        \State $\alpha \gets \alpha_{\mathrm{strong}}$ if $T=\textsc{Up}$, otherwise $\alpha_{\mathrm{small}}$
    \ElsIf{$q > q_{\mathrm{low}}$}
        \State $\alpha \gets \alpha_{\mathrm{strong}}$ if $T=\textsc{Down}$, otherwise $\alpha_{\mathrm{small}}$
    \EndIf
    \If{$\alpha > 0$}
        \State $\hat{B} \gets (1-\alpha)\hat{B}+\alpha B_m$
    \EndIf
\EndIf

\Statex
\State \textbf{Queue-based rate control:}
\If{$q \geq q_{\mathrm{high}}$}
    \If{$s > s_{\mathrm{thres}}$}
        \State $R' \gets R \times M_D$ \Comment{Emergency decrease}
    \ElsIf{$s < -s_{\mathrm{thres}}$}
        \State $R' \gets R$ \Comment{Hold while queue drains}
    \Else
        \State $R' \gets R \times G_D$ \Comment{Gentle decrease}
    \EndIf
\ElsIf{$q_{\mathrm{low}} \leq q < q_{\mathrm{high}}$}
    \State $R' \gets R$ \Comment{Hold under medium queue pressure}
\Else
    \If{$s > s_{\mathrm{thres}}$}
        \State $R' \gets R \times C_I$ \Comment{Cautious increase}
    \Else
        \State $R' \gets R \times A_I$ \Comment{Aggressive probe}
    \EndIf
\EndIf

\Statex
\State \textbf{Apply bandwidth guard:}
\State $R \gets \min(\gamma_{\max}\hat{B},R')$ if the guard is active; otherwise $R \gets R'$
\State $R \gets \max(\gamma_{\min}\hat{B},R)$ if the guard is active and $q<q_{\mathrm{low}}$
\State \Return $R$
\end{algorithmic}
\end{algorithm}

\textbf{Forwarder Congestion Control.}
Algorithm~\ref{alg:congestion-control} shows the rate-control logic used by each forwarder for an upstream face $f$. \system{} first selects the bottleneck feedback between the normalized upstream Interest- and Data-queue occupancies: the larger of $q^{\mathrm{int}}_{\mathrm{face}}(f)$ and $q^{\mathrm{data}}_{\mathrm{face}}(f)$ is used as the queue pressure $q$, and the corresponding slope is used as $s$. This allows the controller to react to the dominant congestion signal regardless of whether pressure appears in the request direction or the data delivery direction.

The controller maintains a smoothed bandwidth estimate $\hat{B}$ using two update rules. If the measured throughput $B_m$ exceeds the current estimate, \system{} treats it as evidence of additional available bandwidth and updates $\hat{B}$ toward $B_m$. If the bottleneck queue pressure exceeds the low threshold, \system{} also updates $\hat{B}$ toward $B_m$ to avoid overestimating usable bandwidth under congestion. The update is stronger when recent throughput samples show a consistent trend, i.e., monotonically increasing for upward probing or monotonically decreasing under queue pressure, and more conservative when the samples fluctuate.

Given the bottleneck queue pressure $q$ and slope $s$, the forwarder then updates the sending rate using a multi-state queue controller. When the queue is high and still increasing, the rate is multiplicatively reduced; when the queue is high but draining, the rate is held; otherwise, the controller applies a gentle decrease. When the queue is in the middle range, the rate is also held.
Only when the queue is low does the forwarder probe for more bandwidth, using a cautious increase if the queue is growing and a more aggressive increase otherwise. Finally, once the bandwidth guard is active, $\gamma_{\max}\hat{B}$ caps the resulting rate, while the $\gamma_{\min}\hat{B}$ floor applies only when $q<q_{\mathrm{low}}$.

\textbf{Consumer Congestion Control.}
Consumers apply the same controller per prefix rather than per face. The Data-queue signal on the application-facing local connection remains zero, so the maximum reduces to the prefix-level Interest-queue feedback $q^{\mathrm{int}}_{\mathrm{flow}}(p)$ and its slope. The prefix is obtained directly from the Interest name, allowing feedback to be mapped to the corresponding flow for per-prefix rate adaptation.

\subsection{Load Balancing}
\label{subsec:load-balancing}

The per-prefix Interest queue in \system{} serves two purposes: shaping the Interest forwarding rate for congestion control and providing a simple substrate for load balancing. After path discovery identifies the valid upstream faces for a prefix, each face maintains an active prefix set, $Prefix_{\text{active}}[f]$, and independently schedules Interests by fetching them from the corresponding per-prefix queues in a round-robin manner. Since ICN follows a receiver-driven request/response model, fair scheduling of Interests generally leads to fair Data delivery as well. \system{} does not rely on caching for load balancing because cache hits are workload-dependent and unpredictable. Instead, it balances Interests directly: each valid upstream face releases Interests according to its congestion-control rate, while round-robin scheduling across per-prefix queues enforces per-prefix fairness and balances utilization across discovered multipaths.

\subsection{\system~ Reliability}
\label{subsec:reliability}

For reliable transfer, \system{} combines per-prefix loss recovery with continued multipath forwarding over available faces.

\system{} sends new Interests only when permitted by both rate pacing and a per-prefix ACK-clocked inflight window $W$; returned Data frees a slot and increases $W$ by $1/W$, while retransmissions remain paced and do not reduce $W$. Over a recent per-prefix RTT sample window, \system{} uses a Jacobson/Karels-inspired estimator~\cite{RTOTimer} and sets $\mathrm{RTO}=\overline{\mathrm{RTT}}+4\sigma_{\mathrm{RTT}}$, where $\overline{\mathrm{RTT}}$ and $\sigma_{\mathrm{RTT}}$ denote the sample mean and standard deviation. To shorten the delivery tail, \system{} triggers fast retransmission for a pending sequence once it has remained unresolved for at least 350\,ms and at least eight later unique Data sequences have been received.

Under loss, ICN's independently named chunks avoid byte-stream HoL blocking, so later Data remain deliverable while a missing chunk is recovered. If a failure makes a face unavailable, a \system{} forwarder continues balancing traffic across the remaining valid faces and resumes using the affected face after recovery.

\section{Implementation}
\label{sec:implementation}

We implement \system{} in \texttt{ndnSIM~2.9}~\cite{ndnsim2.9,ndnsimEvolving} over \texttt{ns-3}~\cite{ns3.19} for simulation and as a Go prototype\footnote{\url{https://github.com/YitongLI2000/MARS-ICNP2026.git}} extending \texttt{ndnd}~\cite{ndnd} with \texttt{ndn-dv} routing~\cite{ndndv} for Mininet emulation. The prototype runs as an NDN-over-UDP overlay and implements \system{} as a stateful forwarding strategy. Its flow-parallel consumer manages each producer prefix independently.

The \system{} forwarder uses a thread-sharded event-loop design. In \texttt{ndnd}, incoming Interests are first dispatched across forwarding threads; \system{} then processes each Interest within the corresponding per-thread strategy instance. Path discovery maintains shared control state because convergence results must be visible across threads, while data transmission keeps per-prefix Interest queues thread-local to avoid fast-path contention. Once a forwarding thread receives and processes a data-transmission Interest, the Interest remains locally owned until transmission, avoiding per-packet migration, preserving locality, and preventing a global queue from becoming the fast-path bottleneck.

Because packet ownership remains local, \system{} coordinates data-transmission scheduling through budgets rather than through a shared packet queue. A dedicated scheduler thread periodically computes token budgets for each upstream face and distributes them across forwarding threads according to local demand. Each forwarding thread consumes its assigned budget locally and drains only its own Interest queues. This avoids races over shared packets, while still allowing \system{} to coordinate multipath rate control at the face level. All emulation variants use application chunks of approximately 6000 bytes, so \system{}'s Interest-count round-robin scheduling approximates byte-level fairness; substantially different chunk sizes would require weighted or deficit scheduling.

\section{Evaluation}
\label{sec:evaluation}

We evaluate \system{} from two complementary perspectives. First, in simulation, we use a broad forwarder deployment in which \system{} runs on participating servers, clients, and core forwarders. This setting stresses the core design under concurrent WAN transfers and tests whether path discovery converges fast, whether it expands usable path diversity, and whether consumer- and forwarder-side congestion control remain tightly coordinated. Second, in Mininet-based prototype emulation, we study more practical deployment assumptions, varying where \system{} forwarders are deployed to understand the performance benefits and overheads of different deployment scales.

\subsection{Evaluation Metrics}
\label{subsec:metric}

We use four primary end-to-end performance metrics. First, \textbf{time to 95\% object delivery ($T_{95}$)} is the elapsed time until 95\% of a flow's object has been delivered and captures near-completion progress. Second, the \textbf{final-5\% tail} is the interval from $T_{95}$ to complete object delivery, complementing $T_{95}$ with the remaining delivery phase to characterize the full object-delivery process. Third, \textbf{per-prefix goodput over time} measures the instantaneous delivered rate for each producer prefix. Since \system{} performs forwarding and control at prefix granularity, this metric exposes utilization, stability, and transient recovery behavior that aggregate throughput may hide. Fourth, \textbf{Jain's fairness index} quantifies balance in average application goodput within each five-flow consumer group. For path discovery, we additionally report usable path exposure, control-packet sizes, startup delay, and discovery time normalized by FCT.

\subsection{Network Topologies}
\label{subsec:network-topo}

\begin{figure}[t]
    \centering
    \includegraphics[width=0.8\columnwidth]{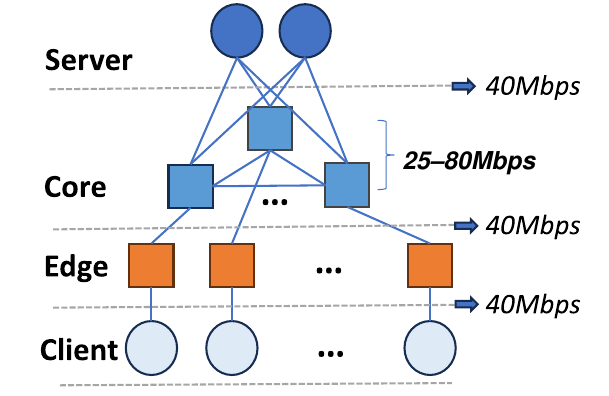}
    \caption{Synthetic hierarchical topology.}
    \label{fig:network-topo}
\end{figure}

Figure~\ref{fig:network-topo} shows the synthetic hierarchical multipath topology used in both evaluation modes. The topology draws on broad WAN design motifs from BT and Google B4, including access/edge/core layering and a connected core~\cite{DINC,B4GloballyDeployedWAN}. For controlled path-exposure experiments, each client, acting as a producer, reaches the core through a single edge branch, whereas each server, acting as a consumer, connects to all five core nodes. This asymmetric attachment pattern creates one shortest route and four longer physical alternatives for each server-client pair, yielding a controlled multipath setting with heterogeneous path lengths.
We evaluate two bandwidth settings over the same topology. In the \textbf{homogeneous} setting, all links are configured with 40~Mbps bandwidth, which isolates algorithmic behavior under uniform link capacities. In the \textbf{heterogeneous} setting, only core-to-core link bandwidths span 25--80~Mbps, while all non-core access/server links remain 40~Mbps, stressing multipath control under unequal WAN bottlenecks. Delays are unchanged across both settings: client/edge-side and server-side access links use 5~ms propagation delay, while core-to-core links use 1~ms propagation delay.

\subsection{Simulation Evaluation}
\label{subsec:simulation}

\subsubsection{Simulation Setup}
\label{subsubsec:simu-setup}

The simulation study focuses on the forwarding-efficiency problem in large-scale ICN-style multipath overlays. Specifically, we use the \textit{homogeneous} setting of the synthetic hierarchical topology in~\secref{subsec:network-topo} to evaluate whether blindly using all routing-provided next hops can degrade forwarding efficiency under concurrent traffic, and whether \system{} can safely expand usable forwarding opportunities while avoiding such inefficient forwarding behavior.

We evaluate a many-to-few data collection workload where multiple servers, acting as consumers, retrieve a 36\,MB object from each geo-distributed client. We vary the scale from 2 to 10 servers and from 20 to 100 clients. To create time-varying contention, different server groups start retrieval at different times, so later transfers must adapt to already occupied WAN capacity.

We compare four systems. \textbf{PCON}~\cite{pcon} represents an ICN forwarder-assisted congestion-control baseline; we use its default forwarding behavior, which can exploit all next hops exposed by the underlying routing system. \textbf{MPTCP}~\cite{mptcp,mptcp-ns3} and \textbf{MPQUIC}~\cite{mpquic-paper} represent end-to-end TCP- and UDP-based multipath transports, respectively; both use ECMP routing, so their path diversity is limited to the routes exposed by the routing layer. \textbf{\system{}} runs its path discovery and forwarder-assisted congestion control over the same WAN topology.

PCON uses its original C++/ndnSIM implementation, while MPTCP and MPQUIC use existing C++/ns-3 implementations.

To stress path discovery and forwarding efficiency at scale, we use a broad \system{} deployment in which \system{} forwarders run on participating servers, clients, and core forwarders. This setting is not intended to model the minimal deployment case; instead, it tests whether \system{} can discover safe forwarding opportunities and maintain coordinated congestion control when many flows concurrently traverse a richly connected WAN overlay.

\subsubsection{Lossless Network}
\label{subsec:evaluation-lossless}

\begin{figure*}[!t]
    \centering

    \begin{minipage}[t]{0.32\textwidth}
        \centering
        \includegraphics[width=\linewidth]{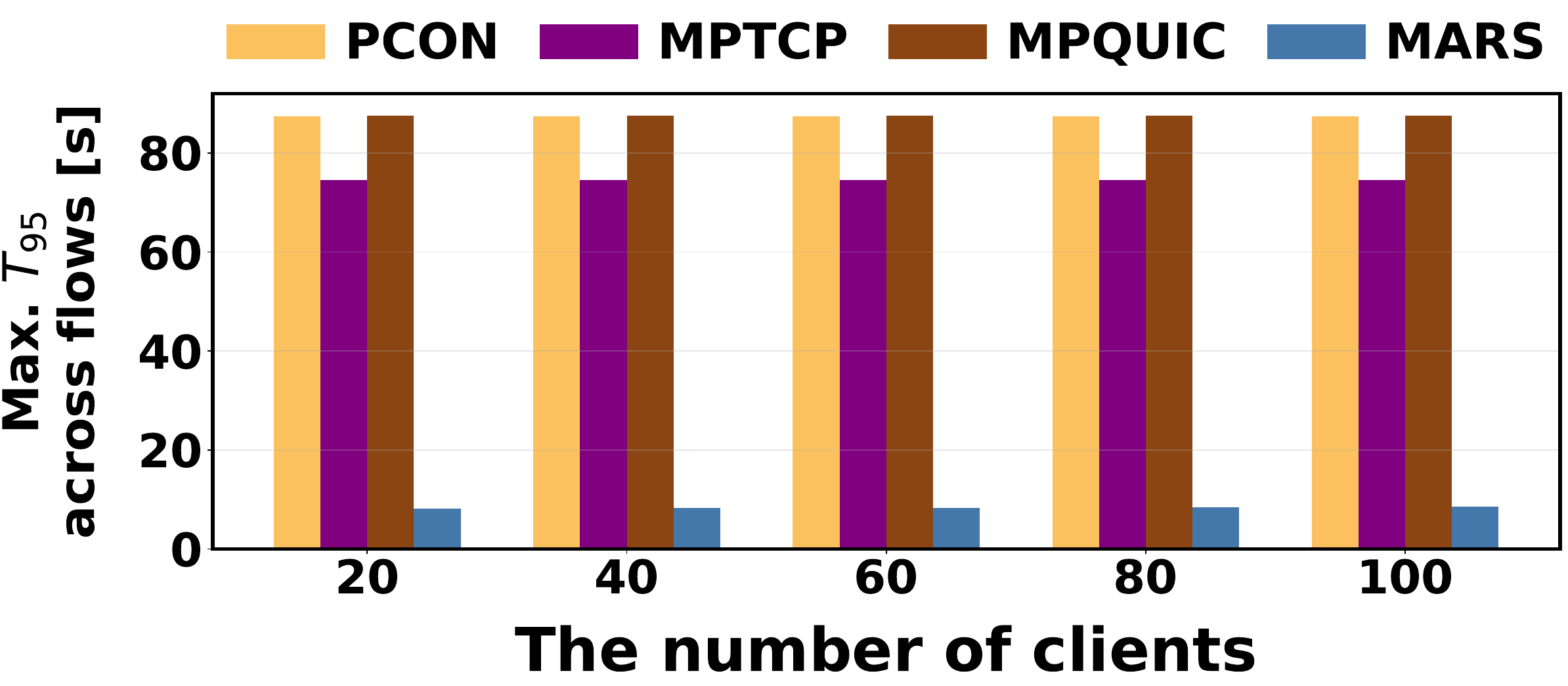}
        \caption{Sim: maximum $T_{95}$ vs. scale.}
        \label{fig:eval_sim_max_fct_async}
    \end{minipage}%
    \hfill%
    \begin{minipage}[t]{0.32\textwidth}
        \centering
        \includegraphics[width=\textwidth]{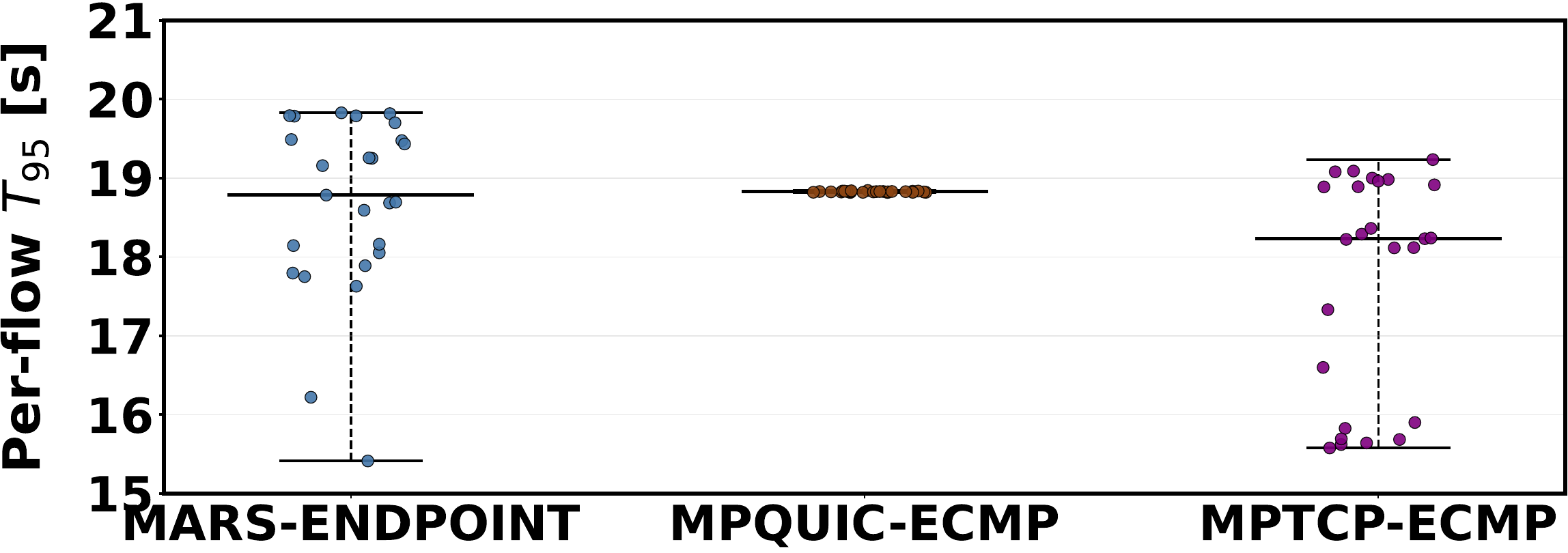}
        \caption{Minimal deployment: lossless per-flow $T_{95}$ distributions.}
        \label{fig:eval_emulation_ablation_lossless_p95}
    \end{minipage}
    \hfill%
    \begin{minipage}[t]{0.32\textwidth}
        \centering
        \includegraphics[width=\textwidth]{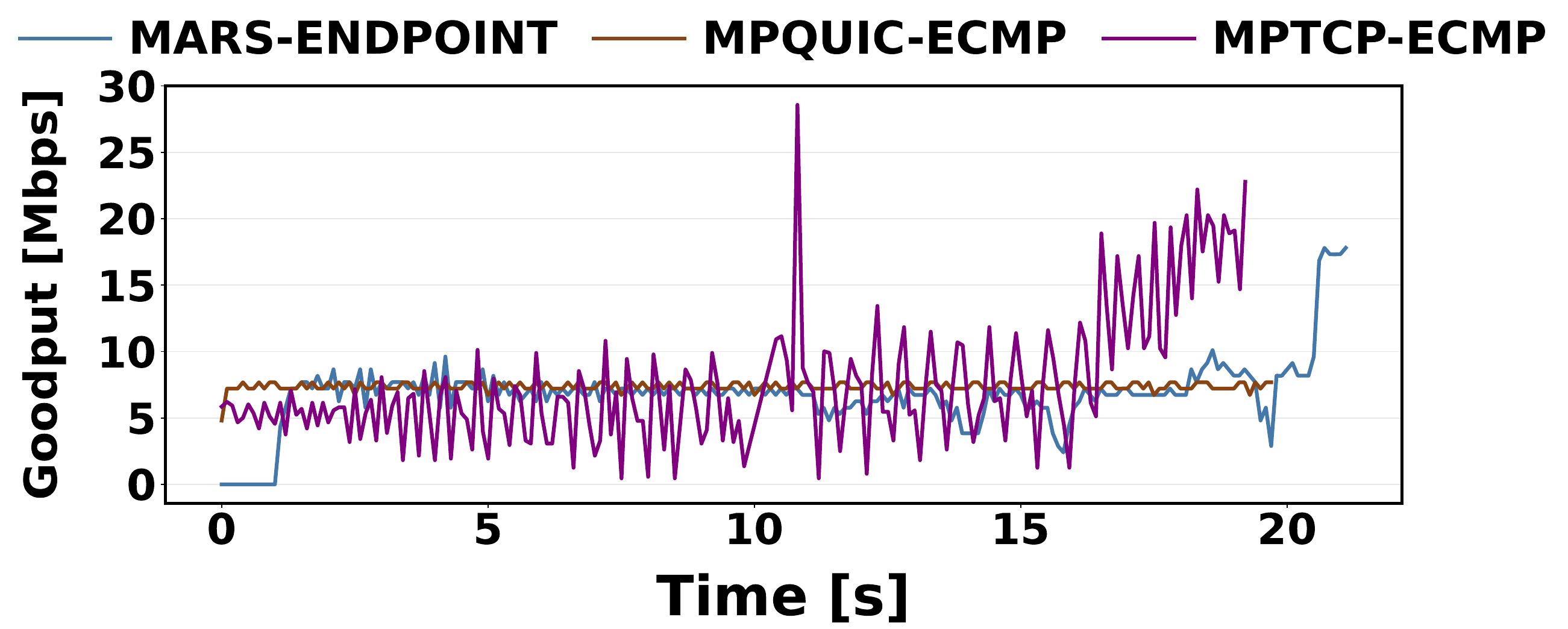}
        \caption{Minimal deployment: goodput.}
        \label{fig:eval_emulation_ablation_lossless_throughput}
    \end{minipage}
\end{figure*}

Fig.~\ref{fig:eval_sim_max_fct_async} shows the maximum $T_{95}$ across flows as the number of clients increases. Across all scales, \system{} keeps this value low and stable, around 8.2--8.6\,s. In contrast, PCON and MPQUIC remain around 87\,s, while MPTCP stays around 75\,s in this broad-deployment stress test, so \system{}'s maximum $T_{95}$ is roughly one order of magnitude lower.
This gap comes from how each system exposes and uses paths. MPTCP and MPQUIC are constrained by ECMP, which exposes only one usable route in this topology, so their nominal multipath schedulers still compete over the same bottleneck path. PCON has the opposite problem: it blindly uses ICN routing-provided next hops, which often creates looped Interests and repeated retransmissions, reducing effective goodput. \system{} instead discovers 10 usable forwarding paths through the core network and filters unsafe next-hop choices, allowing its queue-feedback-based control to shift traffic across safe paths and maintain stable completion time under concurrent load.

\begin{figure*}[!htbp]
    \centering

    \begin{minipage}[t]{0.32\textwidth}
        \centering
        \includegraphics[width=\textwidth]{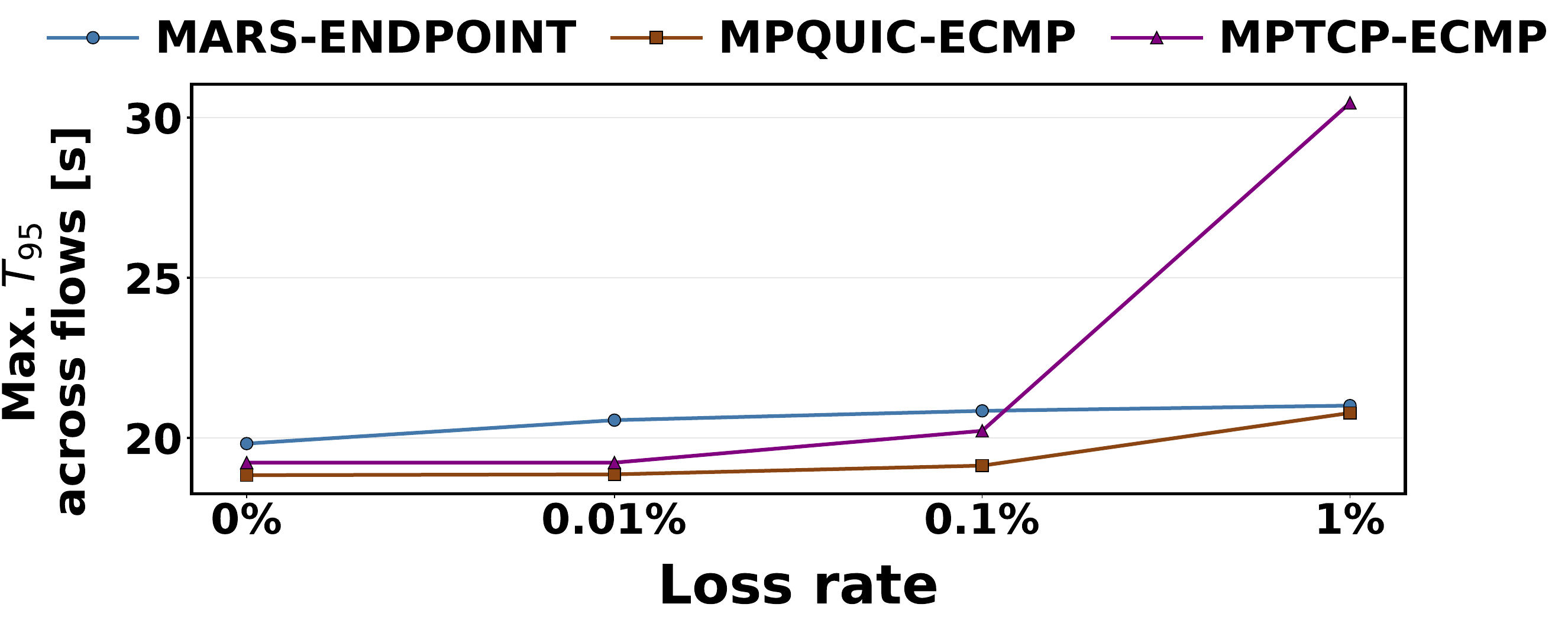}
        \caption{Minimal deployment: maximum $T_{95}$ across flows versus loss rate.}
        \label{fig:eval_emulation_ablation_lossy_p95}
    \end{minipage}
    \hfill
    \begin{minipage}[t]{0.32\textwidth}
        \centering
        \includegraphics[width=\textwidth]{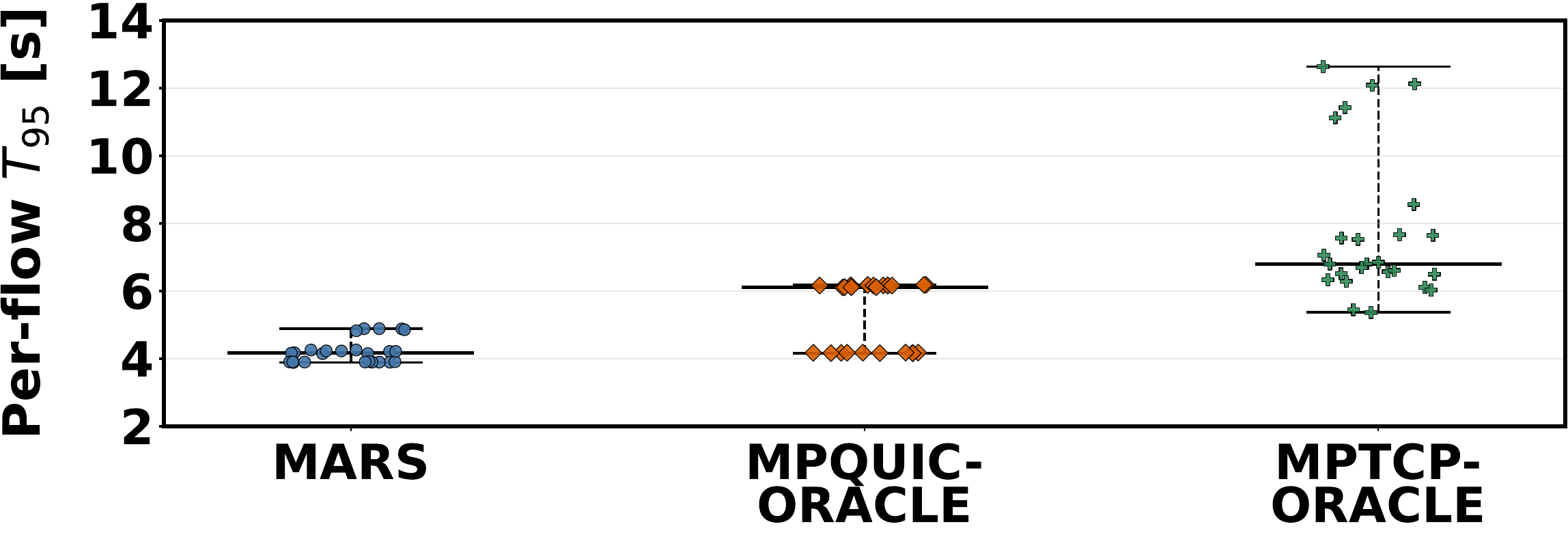}
        \caption{Incremental deployment: lossless per-flow $T_{95}$ distributions.}
        \label{fig:eval_emulation_incremental_lossless_p95}
    \end{minipage}
    \hfill
    \begin{minipage}[t]{0.32\textwidth}
        \centering
        \includegraphics[width=\textwidth]{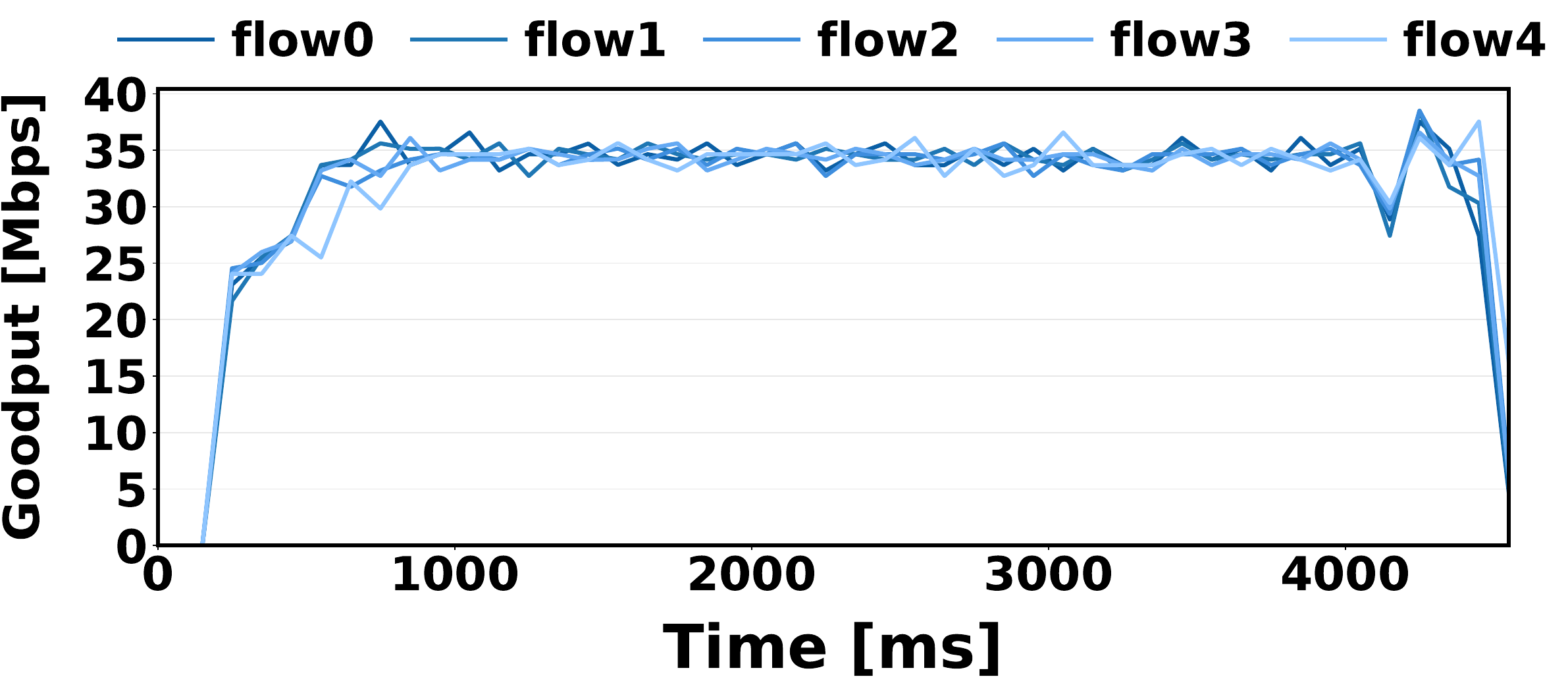}
        \caption{\system{} goodput under incremental deployment.}
        \label{fig:eval_emulation_incremental_lossless_throughput}
    \end{minipage}

\end{figure*}

\begin{figure*}[!htbp]
    \centering
    
    \begin{minipage}[t]{0.32\textwidth}
        \centering
        \includegraphics[width=\linewidth]{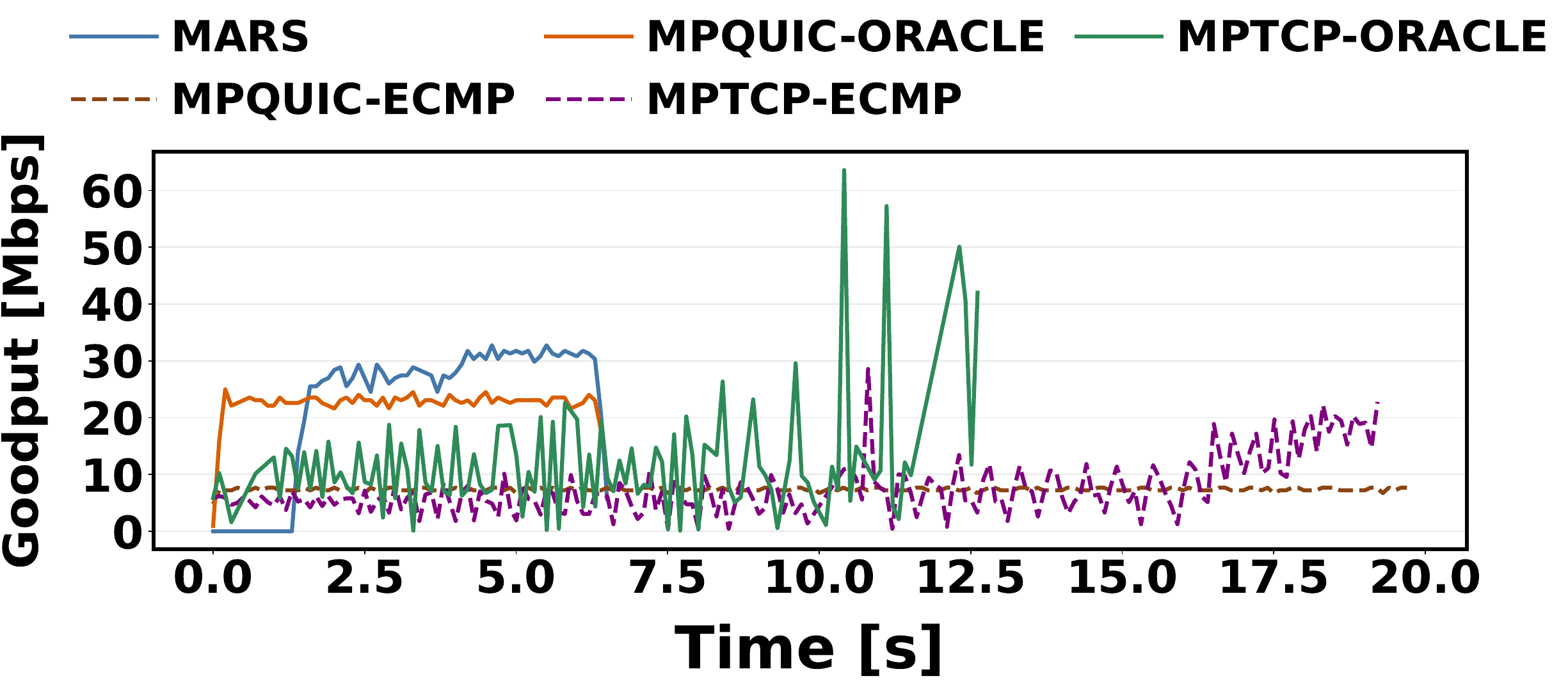}
        \caption{Incremental deployment: goodput comparison.}
        \label{fig:eval_emu_incremental_goodput_all}
    \end{minipage}
    \hfill
    \begin{minipage}[t]{0.32\textwidth}
        \centering
        \includegraphics[width=\textwidth]{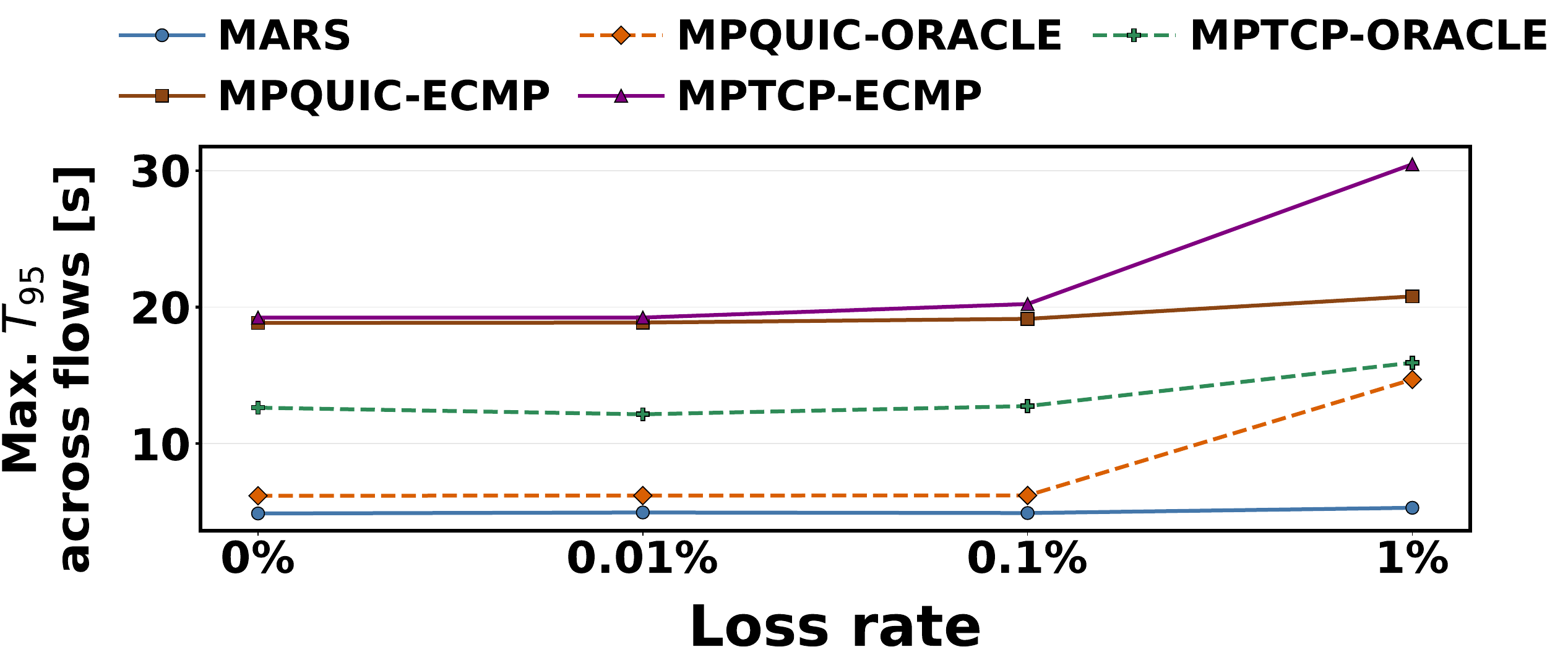}
        \caption{Incremental deployment: maximum $T_{95}$ across flows versus loss rate.}
        \label{fig:eval_emulation_incremental_lossy_p95}
    \end{minipage}
    \hfill    
    \begin{minipage}[t]{0.32\textwidth}
        \centering
        \includegraphics[width=\textwidth]{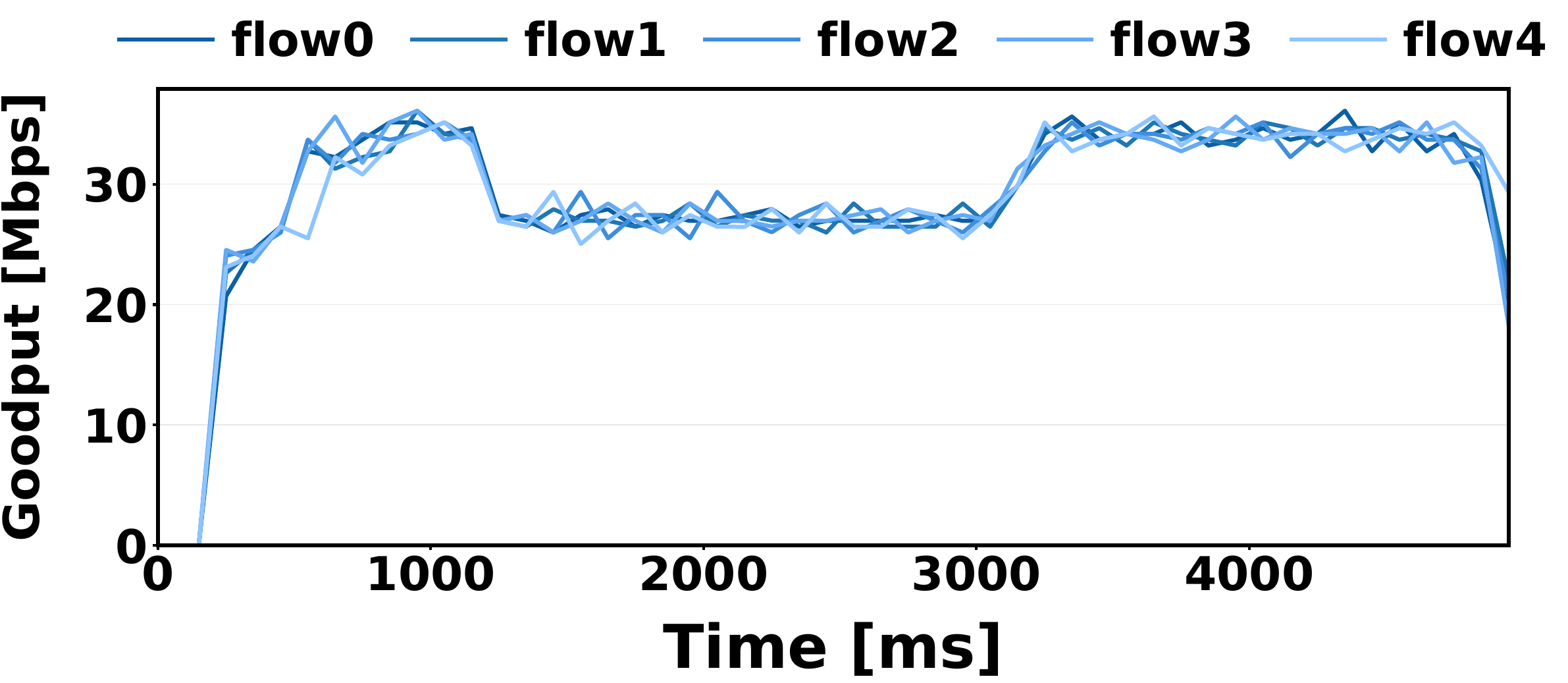}
        \caption{\system{} goodput during forwarding-face outage and recovery.}
        \label{fig:eval_emulation_failure_goodput}
    \end{minipage}

\end{figure*}

We use Shu et al.'s public ns-3 MPQUIC implementation~\cite{mpquic-ns3}, whose application transmits periodically, whereas MPTCP uses a backpressure-driven bulk sender. Shortening MPQUIC's inter-send interval improved performance, and we report its best tested setting. The remaining gap may partly reflect this application-layer traffic-generator difference rather than an inherent protocol difference.

\subsection{Mininet Emulation}
\label{subsec:emulation}

\subsubsection{Emulation Setup}
\label{subsubsec:emulation-setup}

To evaluate the prototype under controlled deployment conditions, we conduct Mininet-based emulation~\cite{mininet} using the prototype described in~\secref{sec:implementation}. The emulation study complements simulation by testing how \system{} behaves under different deployment scopes and by comparing ECMP-limited and path-expanded endpoint configurations with \system{}'s forwarder-assisted design. We evaluate six variants. \textbf{\system{}-ENDPOINT} uses the \system{} consumer and a producer-side \system{} forwarder, while consumer-side and transit forwarders use best-route; without injected consumer-side next hops, it retains one routing-selected path. \textbf{\system{}} additionally enables \system{} at consumer forwarders, allowing path discovery and forwarder-assisted congestion control over the enriched candidate set. \textbf{MPTCP-ECMP} and \textbf{MPQUIC-ECMP} represent controlled ECMP-limited endpoint path-exposure conditions. Because the five physical alternatives have unequal path lengths, the ECMP routing configuration exposes only the shortest route to the endpoint transports.
For the paired \textbf{MPTCP-Oracle} and \textbf{MPQUIC-Oracle} conditions, we deliberately expose all five physical paths of different lengths to MPTCP and MPQUIC through endpoint interfaces, addresses, and routes. Each MPTCP subflow or MPQUIC path is bound to a distinct route using source-address and interface binding with per-path routing tables. These path-expanded Oracle conditions give the endpoint transports the same physical path opportunities as \system{}.

For the evaluated \system{} condition, \texttt{ndn-dv} supplies the initial routing-derived FIB state, which the experiment harness enriches with additional next-hop candidates before path discovery; \system{} then validates five usable paths from this candidate set.

For the baselines, MPQUIC adapts the Go-based \texttt{mp-quic} implementation~\cite{mpquic-paper}, while MPTCP uses the native Linux kernel stack. MPQUIC retains the reference implementation's congestion-control assignment: the initial path uses CUBIC, while additional paths use OLIA. All MPTCP subflows use CUBIC.
Experiments were conducted on one Ubuntu 22.04 server equipped with an Intel Xeon CPU with 128 cores and 256 GB of RAM. We use the same synthetic topology described in~\secref{subsec:network-topo}, with 5 servers and 25 clients to fit the resource constraints of Mininet-based emulation. Unless otherwise stated, emulation uses the \textit{heterogeneous} bandwidth setting. The workload follows a many-to-few collection pattern: each server acts as a receiver/consumer and retrieves an 18\,MB object from each client in its group, forming five concurrent server-client groups. To create time-varying contention, these groups start in a staggered manner at 0\,s, 1\,s, and 2\,s rather than all starting simultaneously.

\subsubsection{Analysis of Path Discovery}
\label{subsec:analysis-pd}

\begin{table}[t]
\centering
\caption{\system{} path-discovery behavior and overhead.}
\label{tab:path-discovery-wan}
\footnotesize
\setlength{\tabcolsep}{3pt}
\begin{tabularx}{\columnwidth}{@{}Xcc@{}}
\toprule
Metric & Lossless & 1\% loss \\
\midrule
\system{}-validated path count & 5 & 5 \\
PD packet sizes (Interest / Data / NACK) & \multicolumn{2}{c@{}}{106 / 107 / 115 bytes} \\
PD startup delay (ms), median [min, max] & 149 [146, 377] & 451 [295, 557] \\
PD time / FCT (separate 36\,MB run) & 1.7\% & 2.8\% \\
\bottomrule
\end{tabularx}
\end{table}

Table~\ref{tab:path-discovery-wan} summarizes \system{} path discovery under the heterogeneous evaluation setting. Under both loss conditions, \system{} validates five usable paths from its FIB candidate set.

The maximum observed PD Interest, Data, and NACK packet sizes are 106, 107, and 115 bytes, respectively. For each five-flow consumer group, PD startup delay is the maximum discovery duration across its flows; the table reports the median [minimum, maximum] across groups. Four of the five groups ran discovery while 5--10 earlier flows were transferring, so the measurement includes workload-induced contention. In a separate 36\,MB run, path-discovery time accounted for 1.7\% and 2.8\% of FCT in the lossless and 1\% loss cases, respectively. Thus, PD completes within 0.56\,s even with background transfers, while its normalized time cost remains below 3\% of FCT in the separate run.

\subsubsection{Minimal Deployment}
\label{subsubsec:minimal-deployment}

We first evaluate a minimal deployment setting, denoted as \textbf{\system{}-ENDPOINT}, using the \system{} consumer and a producer-side \system{} forwarder while consumer-side and transit forwarders use best-route and no extra consumer-side next hops are injected. Its startup PD exchange completes over the routing-selected path but does not expand the path set. Thus, \system{}-ENDPOINT, \textbf{MPQUIC-ECMP}, and \textbf{MPTCP-ECMP} each operate with only one usable path in this experiment. This experiment evaluates \system{} under this minimal deployment in the \textit{heterogeneous} emulation setting.

Fig.~\ref{fig:eval_emulation_ablation_lossless_p95} shows the per-flow $T_{95}$ distributions in the lossless case. \system{}-ENDPOINT spans 15.41--19.83\,s, compared with 18.82--18.84\,s for MPQUIC-ECMP and 15.58--19.23\,s for MPTCP-ECMP. Although the distributions differ, their maximum $T_{95}$ values remain comparable at 19.83\,s, 18.84\,s, and 19.23\,s, respectively. The goodput traces in Fig.~\ref{fig:eval_emulation_ablation_lossless_throughput} show that MPQUIC-ECMP remains nearly constant, \system{}-ENDPOINT is generally stable, and MPTCP-ECMP fluctuates more widely. These results show that even without path-set expansion, \system{} can operate competitively as an endpoint-deployable overlay.

Fig.~\ref{fig:eval_emulation_ablation_lossy_p95} further evaluates robustness under packet loss. From 0\% to 1\% loss, the maximum $T_{95}$ increases gradually from 19.83\,s to 21.01\,s for \system{}-ENDPOINT, compared with 18.84\,s to 20.78\,s for MPQUIC-ECMP and 19.23\,s to 30.46\,s for MPTCP-ECMP. At 1\% loss, \system{}-ENDPOINT remains close to MPQUIC-ECMP and below MPTCP-ECMP. This suggests that \system{}-ENDPOINT remains robust under the evaluated loss rates despite operating with only one usable path.

\subsubsection{Incremental Deployment}
\label{subsubsec:incremental-deployment}

We next evaluate the cooperating client/server deployment in the \textit{heterogeneous} emulation network. We compare \system{} against the ECMP and Oracle baselines defined above. Because the Oracle variants receive the same five paths that \system{} validates from its FIB candidate set, they provide path-opportunity-matched endpoint references.

Fig.~\ref{fig:eval_emulation_incremental_lossless_p95} shows the lossless per-flow $T_{95}$ distribution. \system{} spans 3.89--4.89\,s, while MPQUIC-Oracle spans 4.16--6.18\,s and MPTCP-Oracle spans 5.37--12.63\,s. Thus, \system{} achieves a lower maximum $T_{95}$ than both path-expanded Oracle baselines. Within each endpoint transport, moving from the ECMP-limited to the Oracle configuration reduces the maximum $T_{95}$ from 18.84\,s to 6.18\,s for MPQUIC (67.2\%) and from 19.23\,s to 12.63\,s for MPTCP (34.3\%). These configuration-specific comparisons show the performance benefit of exposing five paths to the endpoint transports, while \system{} further validates and coordinates the same path opportunity through its forwarder-assisted design.

Fig.~\ref{fig:eval_emulation_incremental_lossless_throughput} shows \system{}'s per-flow goodput for one five-flow consumer group. After startup, the five flows remain closely clustered around 34\,Mbps for most of the transfer. Fig.~\ref{fig:eval_emu_incremental_goodput_all} further compares the latest-finishing flow selected from each variant. The \system{} and MPQUIC-Oracle traces sustain higher goodput during their active intervals, while the ECMP traces remain lower and MPTCP-Oracle fluctuates more widely.

Fig.~\ref{fig:eval_emulation_incremental_lossy_p95} evaluates robustness under packet loss. Across loss rates from 0\% to 1\%, \system{} maintains a maximum $T_{95}$ between 4.89\,s and 5.30\,s. At 1\% loss, moving from the ECMP-limited to the Oracle configuration reduces maximum $T_{95}$ by 29.3\% for MPQUIC and 47.7\% for MPTCP. These within-transport reductions quantify the benefit of expanded endpoint path exposure under loss. \system{} maintains the lowest maximum $T_{95}$ across all evaluated loss rates. Since the Oracle variants already expose the same five paths, \system{}'s remaining advantage under loss reflects its coordination of path use through forwarder-assisted feedback and chunk-level retrieval.

\begin{table}[t]
\centering
\caption{Final-5\% per-flow delivery tail under loss (median duration in seconds / median share of total delivery time).}
\label{tab:final-five-tail}
\footnotesize
\setlength{\tabcolsep}{2pt}
\begin{tabularx}{\columnwidth}{@{}Xcccc@{}}
\toprule
& \multicolumn{4}{c}{Packet loss} \\
\cmidrule(l){2-5}
Configuration & 0\% & 0.01\% & 0.1\% & 1\% \\
\midrule
\system{} & 0.37 / 7.53\% & 0.37 / 8.04\% & 0.35 / 7.38\% & 0.81 / 15.45\% \\
MPTCP-Oracle & 0.29 / 3.94\% & 0.36 / 4.66\% & 0.25 / 2.96\% & 0.53 / 4.72\% \\
MPQUIC-Oracle & 0.30 / 4.79\% & 0.30 / 4.76\% & 0.25 / 4.69\% & 0.89 / 6.84\% \\
\bottomrule
\end{tabularx}
\end{table}

To extend the $T_{95}$ analysis through full completion, we define each flow's final-5\% tail as $T_{100}-T_{95}$, where $T_{100}$ is its completion time, and normalize this duration by $T_{100}$; all 25 attempted flows completed in every condition, and Table~\ref{tab:final-five-tail} reports the median of each metric across them. Up to 0.1\% loss, \system{}'s median tail remains 0.35--0.37\,s and accounts for 7.38--8.04\% of total delivery time. At 1\% loss, the median duration and share increase to 0.81\,s and 15.45\%, respectively; the absolute duration lies between MPTCP-Oracle (0.53\,s) and MPQUIC-Oracle (0.89\,s), although its normalized share is higher than both.

\begin{figure}[t]
    \centering
    \includegraphics[width=\columnwidth]{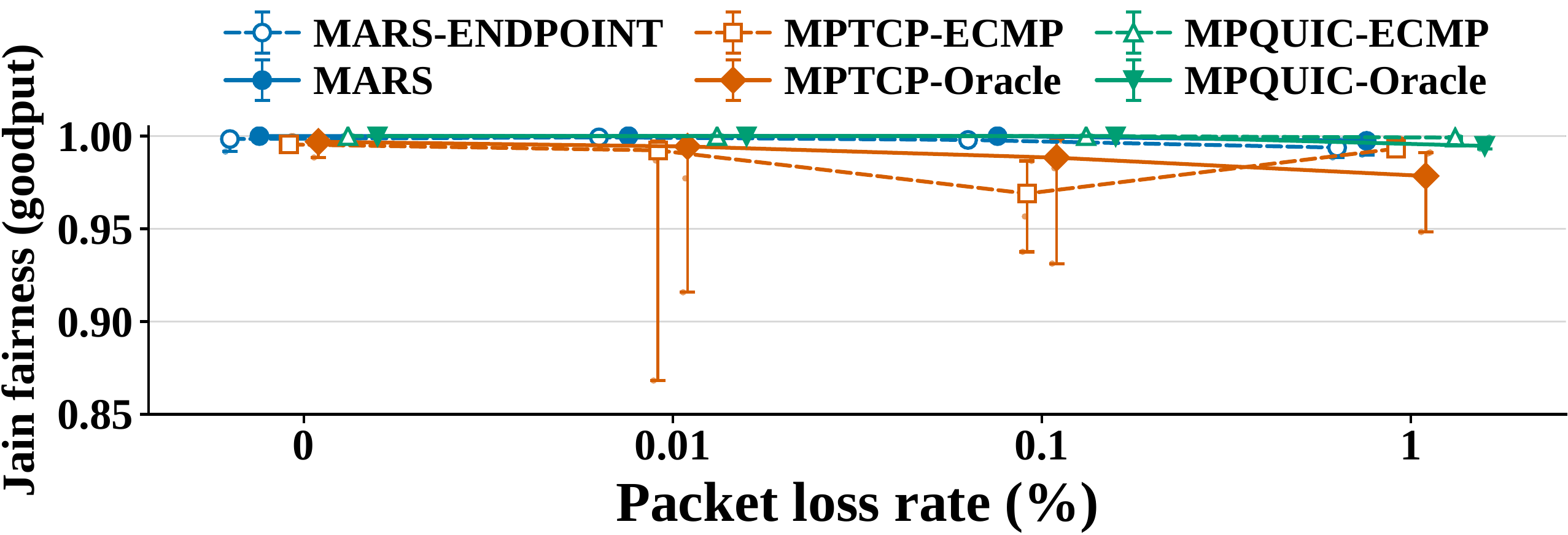}
    \caption{Jain fairness of per-flow average application goodput. Markers show medians across five consumer groups of five flows each; whiskers and small points show the full range and individual groups, respectively.}
    \label{fig:eval_emulation_fairness}
\end{figure}

\subsubsection{Flow Fairness}

We quantify fairness separately within each five-flow consumer group using Jain's index~\cite{jain1984fairness}, $J=(\sum_{i=1}^{n}x_i)^2/(n\sum_{i=1}^{n}x_i^2)$, where $x_i$ is flow $i$'s average application goodput over its complete transfer and $n=5$. For each configuration and loss rate, the five group-level indices describe concurrent scheduling contexts within the same emulation rather than independent repetitions.

Fig.~\ref{fig:eval_emulation_fairness} compares these group-level indices across both deployment settings. Across loss rates from 0\% to 1\%, \system{} maintains a median Jain index of at least 0.99742, with every group at or above 0.98993. \system{}-ENDPOINT's medians range from 0.99378 to 0.99931, while both MPQUIC variants maintain medians of at least 0.99469. MPTCP-ECMP falls to a median of 0.96900 at 0.1\% loss and MPTCP-Oracle to 0.97847 at 1\% loss. Thus, under the evaluated workload, the lower $T_{95}$ achieved by \system{} under incremental deployment is not accompanied by substantial flow-level goodput imbalance.

\FloatBarrier

\subsubsection{Forwarding-Face Outage and Recovery}
\label{subsubsec:failurerecovery}

We evaluate \system{} under an emulated transient forwarding-face outage in the \textit{heterogeneous} setting. For each active consumer, we temporarily exclude one of its five DT-valid forwarding faces from scheduling during the interval from 1\,s to 3\,s after path discovery completes, and then re-enable it. During the outage, \system{} continues forwarding over the remaining discovered faces and resumes using the affected face after restoration.

Relative to normal operation, the forwarding-face outage increases median $T_{95}$ from 4.16\,s to 4.61\,s (10.80\%) and maximum $T_{95}$ from 4.89\,s to 5.17\,s (5.77\%). Across the stable pre-outage and outage intervals in Fig.~\ref{fig:eval_emulation_failure_goodput}, the mean per-flow goodput of the five displayed flows drops by 19.41\%; it returns to at least 95\% of its pre-outage level approximately 250\,ms after face restoration.

\subsection{Discussion: Deployment Scope and Limitations}
\label{subsec:discussion}

\system{} targets service-managed or cooperative overlays, such as CDNs, edge/cloud platforms, and application-managed relays, rather than replacing Internet-wide interdomain routing. Its path-discovery benefit depends on forwarder placement and the candidate set exposed by the local FIB: endpoint-only or sparse deployments may expose little beyond ECMP, whereas broader deployment can combine safe path validation with congestion feedback closer to bottlenecks.

Our evaluation uses controlled synthetic topologies and a fixed large-object collection workload. Validation on larger multi-host testbeds and operational overlays, with more diverse workloads, routing policies, traffic dynamics, and forwarder placements, remains future work. \system{} also exhibits a higher median final-5\% tail share than both path-expanded Oracle baselines across the evaluated loss rates, with the largest gap at 1\% loss. Finally, the baseline results are configuration-specific: MPTCP uses CUBIC on all subflows, whereas MPQUIC retains its released CUBIC/OLIA assignment; alternative configurations, such as BBR for MPTCP~\cite{bbr} or all-CUBIC for MPQUIC, are not evaluated.

\section{Related Work}
\label{sec:related-work}

\subsection{End-to-End Multipath Transports}
\label{subsec:e2e-multipath}

End-to-end multipath transports exploit endpoint-visible path diversity under a single transport-layer abstraction, allowing applications to use multiple paths without directly managing scheduling, reliability, or congestion control. CMT-SCTP extends SCTP~\cite{rfc9260} multihoming from failover to concurrent data transfer~\cite{cmt-sctp}. MPTCP maintains multiple TCP subflows under one reliable byte-stream connection~\cite{rfc8684}, while MPQUIC brings this model to QUIC using its UDP-based, user-space design and connection identifiers~\cite{mpquic-paper,mpQUIC}. These designs are deployable at endpoints, but their path diversity is limited to addresses, interfaces, subflows, or path identifiers exposed by the underlying network.

\subsection{Routing-Level and Path-Aware Multipath}
\label{subsec:routing-multipath}

Routing-level and path-aware approaches expose path diversity through the network control plane. ECMP distributes traffic across equal-cost next hops using flow hashing~\cite{ECMP}, while interdomain multipath systems such as YAMR construct alternate policy-compliant paths from BGP's default route for resilience and traffic distribution~\cite{yamr}. Path-aware networking further exposes path properties or choices to endpoints and applications, enabling path selection based on performance, resilience, or policy objectives~\cite{schmitt2018pathaware}. These approaches can provide richer path visibility than endpoint-only transports, but generally require router, routing-protocol, domain-level, or operator support, making incremental Internet/WAN deployment difficult.

\subsection{Overlay- and Forwarder-Assisted Multipath}
\label{subsec:overlay-multipath}

Overlay- and forwarder-assisted approaches introduce intermediate nodes above IP routing to expose additional forwarding choices. RON and Overlay TCP use application-layer routers to route around degraded paths or support multipath routing and congestion control~\cite{ron,overlaytcp}. In ICN, PCON, MIRCC, and NDN-QSF exploit stateful forwarding and congestion feedback to steer Interests across routing-provided next hops~\cite{pcon,mircc,ndnqsf}. However, these systems typically take routing-provided next hops as given, rather than discovering and filtering safe overlay forwarding choices.
Recent proxying mechanisms such as MASQUE/CONNECT-UDP further show that UDP-based overlay forwarding is deployable in practice~\cite{rfc9298}. Together, these systems show the value of adaptive forwarding and deployable UDP overlays, but they do not jointly coordinate safe path discovery, endpoint scheduling, congestion response, and intermediate forwarding across cooperating forwarders.

\section{Conclusion}
\label{sec:conclusion}

This paper presented \system{}, a receiver-driven, forwarder-assisted multipath transport for Internet/WAN environments. \system{} combines tier-synchronized overlay path discovery with coupled consumer/forwarder congestion control to safely expand usable forwarding opportunities and coordinate traffic across heterogeneous paths. Unlike routing-level multipath approaches, \system{} does not require changes to IP routers, routing protocols, or interdomain control planes; instead, it runs as a UDP-based overlay that can be incrementally deployed at clients, servers, relays, or service-managed edge nodes. Our simulation and Mininet prototype evaluation show that this modest overlay coordination can improve multipath delivery under different deployment scales, loss conditions, and the evaluated forwarding-face outage. Overall, \system{} demonstrates that service-managed forwarders can provide a practical extension point between endpoint-only transport and routing-level multipath, enabling richer path use and faster congestion response while preserving compatibility with today’s Internet infrastructure.

\clearpage

\IEEEtriggeratref{40}
\bibliographystyle{IEEEtran}
\bibliography{reference}

\end{document}